\documentclass [11pt]{article}%
\usepackage{epsf} 
\usepackage{amsmath}
\usepackage{amssymb}
\usepackage{epsfig}
\usepackage{latexsym}
\usepackage{amsfonts}
\usepackage{graphicx}%
\usepackage{varioref}
\usepackage{ifthen}
\begin{document}
\parindent 0mm 
\setlength{\parskip}{\baselineskip} 
\thispagestyle{empty}
\pagenumbering{arabic} 
\setcounter{page}{0}
\mbox{ }
\rightline{UCT-TP-283/10}
\newline
\rightline{October 2010}
\newline
\rightline{Revised: February 2011}
\newline
\vspace{0.2cm}
\begin{center}
{\Large {\bf (Pseudo)Scalar Charmonium in   Finite Temperature QCD}}
{\LARGE \footnote{{\LARGE {\footnotesize Supported in part by FONDECYT 1095217  (Chile),  Proyecto Anillos ACT119 (Chile), and NRF (South Africa).}}}}
\end{center}
\vspace{.1cm}
\begin{center}
{\bf C. A. Dominguez}$^{(a),(b)}$,  {\bf M. Loewe},$^{(c)}$, {\bf J. C. Rojas}$^{(d)}$,
 {\bf Y. Zhang}$^{(a)}$
\end{center}
\begin{center}
$^{(a)}$Centre for Theoretical \& Mathematical Physics,University of
Cape Town, Rondebosch 7700, South Africa

$^{(b)}$ Department of Physics, Stellenbosch University, Stellenbosch 7600, South Africa

$^{(c)}$Facultad de F\'{i}sica, Pontificia Universidad Cat\'{o}lica de Chile, Casilla 306, Santiago 22, Chile

$^{(d)}$ Departamento de F\'{i}sica,  Universidad Cat\'{o}lica del Norte, Casilla 1280, Antofagasta, Chile\end{center}
\vspace{0.3cm}
\begin{center}
\textbf{Abstract}
\end{center}
The hadronic parameters of pseudoscalar ($\eta_c$) and scalar  ($\chi_c$) charmonium are determined at finite temperature from  Hilbert moment QCD sum rules. These parameters are the hadron mass, leptonic decay constant, total width, and continuum threshold ($s_0$). Results for $s_0(T)$ in both channels indicate that $s_0(T)$ starts approximately constant, and then it decreases monotonically with increasing $T$ until it reaches the QCD threshold, $s_{th} = 4 m_Q^2$, at a critical temperature $T = T_c \simeq  180 \; \mbox{MeV}$ interpreted as the deconfinement temperature. 
The other hadronic parameters behave qualitatively similarly to those of the $J/\psi$, as determined in this same framework. The hadron mass is essentially constant, the total width is initially independent of T, and after $T/T_c \simeq 0.80$ it begins to increase with increasing $T$ up to $T/T_c  \simeq 0.90 \; (0.95)$ for $\chi_c$ ($\eta_c$), and subsequently it decreases sharply up to $T \simeq 0.94 \; (0.99) \; T_c$, for $\chi_c$ ($\eta_c$), beyond which the sum rules are no longer valid.
The decay constant of $\chi_c$ at first remains basically flat  up to $T \simeq 0.80\; T_c$, then it starts to decrease up to $T \simeq 0.90 \;T_c$, and finally it increases  sharply with increasing $T$.
In the case of $\eta_c$ the decay constant does not change up to  $T \simeq 0.80 \;T_c$ where it begins a gentle increase up to $T \simeq 0.95 \;T_c$  beyond which it increases dramatically with increasing $T$. This behaviour contrasts with that of light-light and heavy-light quark systems, and it  suggests the survival of the $\eta_c$ and the $\chi_c$ states beyond the critical temperature, as already found for the $J/\psi$ from similar QCD sum rules. These conclusions are very stable against changes in the critical temperature in the wide range $T_c = 180 -  260 \; \mbox{MeV}$.\\ 

\newpage
\bigskip
\noindent
\section{INTRODUCTION}
\noindent
The extension of the QCD sum rule method \cite{REV} to finite temperature was first discussed in \cite{BOCH}, and subsequently developed successfully to obtain the $T$-dependence of hadronic parameters, and study the phase transitions of chiral-symmetry restoration and quark gluon deconfinement \cite{CAD1} - \cite{JPSI}. Further evidence for the validity of this extension to finite temperature was given in \cite{DL95}.
It has been established in this framework that hadronic spectral functions experience substantial modifications with increasing $T$. In particular, in the case of light-light and heavy-light quark systems, their leptonic decay constants, weak and strong couplings,  hadronic widths, and form factors develop a  prominent $T$-dependence consistent with deconfinement at some critical temperature $T_c$. The advocacy of a substantial temperature dependence of hadronic masses has been refuted by results from  these applications. In fact, these masses hardly change with $T$. If $s_0$ is the squared energy beyond which the hadronic spectral function merges into a continuum, well represented by perturbative QCD (PQCD), it has been found that $s_0(T)/s_0(0) \simeq \langle 0| \bar{q} q | 0 \rangle|_T / \langle 0| \bar{q} q | 0 \rangle|_{T=0}$ \cite{CAD1}. The vanishing of the quark condensate signals chiral symmetry restoration, while the vanishing of the continuum threshold $s_0(T)$ is an indication of quark-gluon deconfinement, as resonances disappear and the smooth spectral function is determined purely by PQCD. The gluon condensate plays essentially no role here due to its peculiar behaviour, i.e. it hardly changes until $T$ becomes very close to $T_c$. However, this will not be the case for heavy-heavy quark systems due to the absence of the light-quark condensate. Hence, a qualitatively different temperature behaviour of charmonium is to be expected. In fact, we have shown recently \cite{JPSI} that the $J/\psi$  total width initially increases with increasing $T$ up to some temperature $T^* < T_c$ after which it begins to decrease beyond $T_c$, while the leptonic coupling remains approximately constant up to $T^*$, and subsequently it increases with temperature also beyond $T_c$. This behaviour provides strong evidence for the survival of charmonium in the vector channel beyond the critical temperature $T_c$, in agreement with lattice QCD results \cite{lattice}. The $T$ dependence of $s_0(T)$ in the $J/\psi$ channel is qualitatively similar to that for light-light and heavy-light quark systems, i.e. $s_0(T)$ decreases monotonically with increasing $T$ until it reaches the QCD threshold $s_{th} = 4 m_Q^2$, with $m_Q$ being the quark mass. Beyond this point there is no longer a solution to the FESR as the integration range vanishes.
The dynamical origin of this behaviour lies with the gluon condensate as well as with the $T$-dependence of the PQCD vector correlator in the space-like region, the so called current-quark scattering term. In fact, there is a distinct interplay between these two terms with increasing temperature.\\
At this stage it must be pointed out that in the framework of QCD sum rules the critical temperature for deconfinement, referred to above, is just a phenomenological parameter. It is the temperature at which the resonance couplings and the continuum threshold approach zero, and the widths increase sharply, for light-light and heavy-light quark correlators. Hence, it need not coincide numerically with e.g. the critical temperature obtained in lattice QCD, which is defined differently. In fact, results from QCD sum rules lead to  values of $T_c$ somewhat different from those from lattice QCD. Hence, comparisons between different frameworks should be made in terms of the dependence of parameters on the ratio $T/T_c$.\\
In this paper we extend the analysis of \cite{JPSI} to charmonium in the scalar ($\chi_c$) and the pseudoscalar ($\eta_c$) channels. We find a temperature behaviour qualitatively similar to that of $J/\psi$, i.e. a monotonically decreasing $s_0(T)$, suggesting deconfinement at a critical temperature, a total width which initially is independent of $T$, and then it increases with $T$ up to  $T/T_c \simeq 0.90 - 0.95$,  beyond which it begins to decrease dramatically, and a leptonic coupling which starts constant, then begins to decrease  for $\chi_c$ (or already starts to increase for the $\eta_c$), and finally shoots up sharply in both cases. The hadronic masses remain essentially constant. In these two channels there is no contribution  to the correlator in the space-like region (scattering term), so that the gluon condensate is responsible for all of the thermal behaviour of these hadronic states. Due to this circumstance, the critical temperature for deconfinement, $T_c$,  corresponds to that at which the gluon condensate vanishes. Given the uncertainty in the value of $T_c$ we consider the wide range
$T_c \simeq \; 180 - 260 \; \mbox{MeV}$. Qualitatively, the results remain very stable in this range. Slight quantitative variations, however, do not affect the conclusions.
Since $s_0(T)$ approaches the QCD threshold $s_{th} = 4 m_Q^2$ as $T$ approaches $T_c$, the integration range in the FESR approaches zero, and the method is no longer valid. Nevertheless,  the dramatic decrease of the width, and the increase of the decay constant near $T_c$ suggests the survival of the $\eta_c$ and $\chi_c$ beyond the critical temperature for deconfinement, in agreement with lattice QCD results \cite{lattice}.

\section{ QCD SUM RULES}
\noindent
We consider the correlator of the heavy-heavy quark (pseudo)scalar current at finite temperature

\begin{equation}
\Pi_{(5)} (q^{2},T)   = i \, \int\; d^{4} \, x \; e^{i q x} \; \;\theta(x_0)\;
<<|[ J_{(5)}(x) \;, \; J_{(5)}^{\dagger}(0)]|>> \;,
\end{equation}

where $J_{(5)}(x) = : \bar{Q}(x) \Gamma Q(x):$, $\Gamma =\gamma_5$ or $\Gamma = 1$ for the (pseudo)scalar case,  and $Q(x)$ is the heavy quark field.
The vacuum to vacuum matrix element above is the Gibbs average

\begin{equation}
<< A \cdot B>> = \sum_n exp(-E_n/T) <n| A \cdot B|n> / Tr (exp(-H/T)) \;,
\end{equation}

where $|n>$ is any complete set of eigenstates of the (QCD) Hamiltonian. We shall adopt the quark-gluon basis, as this allows for the standard QCD sum rule program at $T=0$ to be seamlessly extended to $T \neq 0$ \cite{CAD2}. The QCD sum rules to be used are the  Hilbert moments

\begin{equation}
\varphi_N(Q_0^2,T) \equiv \frac{1}{(N+1)!}\, \Bigl(- \frac{d}{dQ^2}\Bigr)^{(N+1)} \Pi_{(5)}(Q^2,T)|_{Q^2=Q_0^2} = \frac{1}{\pi}
\int_{0}^{\infty} \frac{ds}{(s+Q_0^2)^{(N+2)}}\,  Im \,\Pi(s,T)\; , 
\end{equation}

where $N = 1,2,...$, and $Q_0^2 \geq 0$ is an external four-momentum squared to be considered as a free parameter \cite{RRYNP}. 
Using Cauchy's theorem in the complex s-plane, which is equivalent to invoking quark-hadron duality, the Hilbert moments become Finite Energy QCD sum rules (FESR), i.e.

\begin{equation}
\varphi_N(Q_0^2, T)|_{HAD} =  \varphi_N(Q_0^2,T)|_{QCD} \;,
\end{equation}

where
\begin{equation}
\varphi_N(Q_0^2,T)|_{HAD} \equiv \frac{1}{\pi}
\int_{0}^{s_0(T)}\frac{ds}{(s+Q_0^2)^{(N+2)}}\, Im \,\Pi(s,T)|_{HAD} \;,
\end{equation}

\begin{equation}
 \varphi_N(Q_0^2,T)|_{QCD} \equiv \frac{1}{\pi}
\int_{4 m_Q^2}^{s_0(T)}\frac{ds}{(s+Q_0^2)^{(N+2)}}\, Im \,\Pi_{PQCD}(s,T) 
+ \varphi_N(Q_0^2,T)|_{NP}  \;.
\end{equation}

The imaginary part of the (pseudo)scalar correlator  in perturbative QCD (PQCD) at finite temperature, $\mbox{Im} \; \Pi_{(5)}(q^2,T)$, involves two pieces, one in the time-like region ($q^2 \geq 4 m_Q^2$), $\mbox{Im} \; \Pi_{(5)}^{(a)}(q^2,T)$, which survives at T=0, and one in the space-like region ($q^2 \leq 0$), $\mbox{Im} \; \Pi_{(5)}^{(s)}(q^2,T)$, which vanishes at T=0. A straightforward calculation of the pseudoscalar correlator in the time-like region, at $T \neq 0$, and to leading order in PQCD, gives 

\begin{equation}
\frac{1}{\pi}\, Im \,\Pi_5^{(a)}(\omega,T) =  \frac{3}{8 \pi^2}\; \omega^2 \; v(\omega^2) \left[ 1 - 2 \, n_F\left(\frac{\omega}{2T}\right)\right] \theta(\omega^2 - 4 m_Q^2) \;,
\end{equation}

where $ [v(\omega^2)]^2 = 1 - 4 m_Q^2/\omega^2$, $m_Q$ is the heavy quark mass, $q^2 = \omega^2 - \mathbf{q}^2 = \omega^2$ in the rest frame of the thermal bath, and  $n_F(z) = (1+e^z)^{-1}$ is the Fermi thermal function.
The quark mass is assumed independent of $T$, which is a good approximation for temperatures below 200-250 MeV \cite{mQ}. The equivalent expression for the scalar correlator is

\begin{equation}\frac{1}{\pi}\, Im \,\Pi^{(a)}(\omega,T) =  \frac{3}{8 \pi^2}\; \omega^2 \; [v(\omega^2)]^3 \left[ 1 - 2 \, n_F\left(\frac{\omega}{2T}\right)\right] \theta(\omega^2 - 4 m_Q^2) \;.
\end{equation}

Both spectral functions are strongly exponentially suppressed. In the so called scattering region ($q^2 < 0$) the scalar QCD spectral function is given by

\begin{eqnarray}
\frac{1}{\pi}\, Im \,\Pi^{(s)}(q^2,T) &=&  \frac{9}{16 \pi^2}\; q^2 \int_{v}^{\infty} x^2\; \left[n_F\left(\frac{|{\bf q}| x + \omega}{2 T}\right) - n_F\left(\frac{|{\bf q}| x - \omega}{2 T}\right) \right] \; dx
\nonumber \\ [.3cm]
&=& \frac{9}{2 \pi^2}\,  T^2\, \left(\frac{\omega}{|{\bf q}|^3}\right) \, q^2 \int_{\frac{|{\bf q}| v}{ 2T}}^{\infty} z^2\, \frac{d}{dz} n_F(z) \;dz \;.
\end{eqnarray}
Integrating by parts, the above equation becomes

\begin{equation}
\frac{1}{\pi}\, Im \,\Pi^{(s)}(q^2,T) = - \;\frac{9}{2 \pi^2}\; q^2\;T^2 \left(\frac{\omega}{|{\bf q}|^3}\right) \left[\frac{m_Q^2}{T^2}\; n_F\left(\frac{m_Q}{T}\right) + 2 \int_{\frac{m_Q}{T}}^{\infty} z \, n_F(z) \,dz \right] \;,
\end{equation}

where  $\lim_{\stackrel{|\bf{q}| \rightarrow 0}{\omega \rightarrow 0}}\; |{\bf q}|\; v = 2\; m_Q$ has been used above. Invoking the limit

\begin{equation}
\lim_{\stackrel{|\bf{q}| \rightarrow 0}{\omega \rightarrow 0}}\;\frac{\omega}{|{\bf q}|^3} = \frac{2}{3}\; \delta(\omega^2) \;,
\end{equation}

the QCD spectral function, Eq.(10), vanishes since $\lim_{\stackrel{|\bf{q}| \rightarrow 0}{\omega \rightarrow 0}} \;\frac{\omega}{|{\bf q}|^3} \;q^2 = \lim_{\stackrel{|\bf{q}| \rightarrow 0}{\omega \rightarrow 0}}\; \frac{2}{3}\; (\omega^2 - |{\bf q}|^2) \;\delta(\omega^2) = \frac{2}{3}\; \omega^2 \delta(\omega^2) = 0.$ For the pseudoscalar correlator the QCD spectral function in the scattering region is

\begin{equation}
\frac{1}{\pi}\, Im \,\Pi_5^{(s)}(q^2,T) =  \frac{3}{8 \pi^2}\; q^2 \int_{v}^{\infty} \; \left[n_F\left(\frac{|{\bf q}| x + \omega}{2 T}\right) - n_F\left(\frac{|{\bf q}| x - \omega}{2 T}\right) \right] \; dx
 \;.
\end{equation}

In the rest frame of the thermal bath this contribution vanishes, i.e.

\begin{equation}
\lim_{\stackrel{|\bf{q}| \rightarrow 0}{\omega \rightarrow 0}}\;\frac{1}{\pi} \, Im \,\Pi_5^{(s)}(q^2,T) = - \, \frac{3}{4 \pi^2}\; q^2\; \left(\frac{\omega}{|{\bf q}|}\right) \; n_F\left(\frac{\omega}{T}\right) \rightarrow 0\;.
\end{equation}

\begin{figure}[ht]
\begin{center}
\includegraphics[width=\columnwidth]{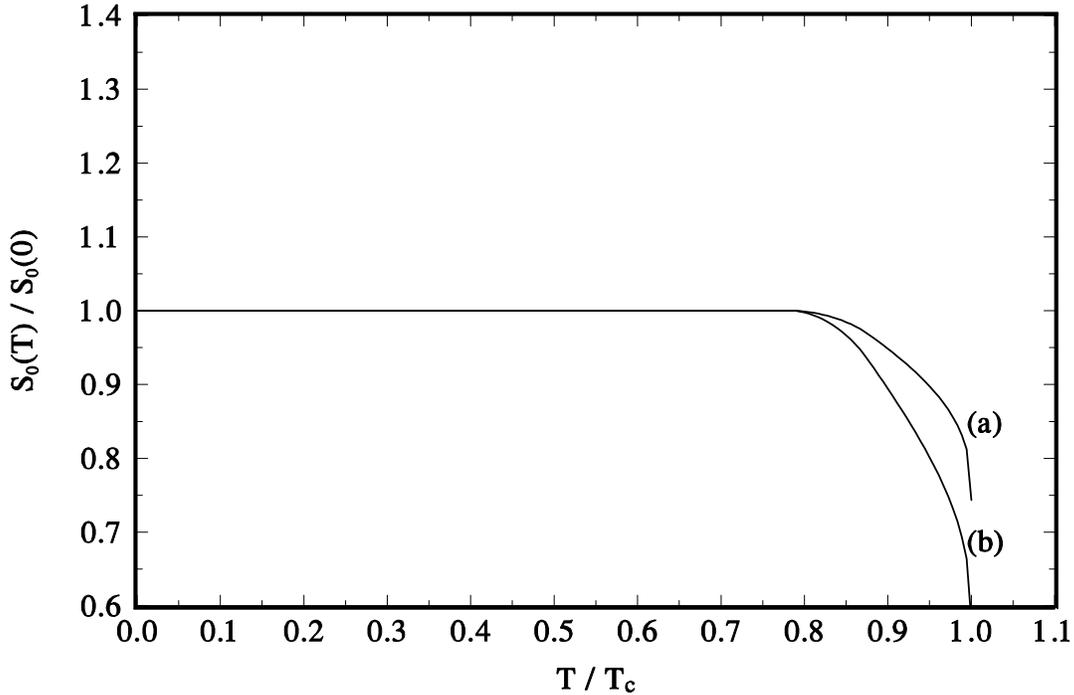}
\caption{The  ratio $s_0(T)/s_0(0)$  as a function of $T/T_c$, for $\eta_c$ (curve (a)), and $\chi_c$ (curve (b)) corresponding to $T_c = 180 \; \mbox{MeV}$. For $T_c = 260 \; \mbox{MeV}$ the break-point is at $T/T_c \simeq 0.55$, instead of $T/T_c \simeq 0.80$.}
\end{center}
\end{figure}

This leaves the gluon condensate as the leading temperature dependent term in the OPE. The moments for this term have been calculated at $T=0$ in \cite{RRYNP}. After taking into account that our definition, Eq.(3), differs from that in  \cite{RRYNP} by one additional derivative, the results for the scalar correlator are

\begin{eqnarray}
\varphi_{N}(Q^2,T)|_{NP} &=& - \frac{27}{8\pi^2}\frac{2^{(N+1)} N !}{(4m_Q^2)^{(N+1)}}\frac{1}{(1+\xi)^{N+2}}
\frac{(N+1)(N+2)(N+3)(N+4)}{(2N+5)!!\;(2N + 7)} \nonumber \\ [.3cm]
&\times& 
\left[F\left(N+2,-\frac{1}{2},N+\frac{9}{2};\rho\right)
-
\frac{2}{3 (N+4)} F\left(N+2,\frac{1}{2},N+\frac{9}{2};\rho\right)\right] \Phi \,,
\end{eqnarray}

where $F(a,b,c,z)$ is the hypergeometric function, , $\xi\equiv\frac{Q_0^2}{4m_Q^2}$, $\rho\equiv\frac{\xi}{1+\xi}$, and

\begin{equation}
\Phi\equiv\frac{4\pi^2}{9}\frac{1}{(4m_Q^2)^2}\left<\left<\frac{\alpha_s}{\pi}G^2\right>\right>\;,
\end{equation}

where $\left<\left<\frac{\alpha_s}{\pi}G^2\right>\right>$ stands for the temperature dependent gluon condensate.
The result for the pseudoscalar is

\begin{eqnarray}
\varphi_{5 N}(Q^2,T)|_{NP} &=& - \frac{3}{8\pi^2}\frac{2^{(N+1)} N !}{(4m_Q^2)^{(N+1)}}\frac{1}{(1+\xi)^{N+2}}
\frac{(N+1)(N+2)(N+3)(N+4)}{(2N+3)!!\;(2N + 4)} \nonumber \\ [.3cm]
&\times& 
\left[F\left(N+2,-\frac{3}{2},N+\frac{7}{2};\rho\right)
-
\frac{6}{(N+4)} F\left(N+2,-\frac{1}{2},N+\frac{7}{2};\rho\right)\right] \Phi \,.
\end{eqnarray}

\begin{figure}[ht]
\begin{center}
\includegraphics[width=\columnwidth]{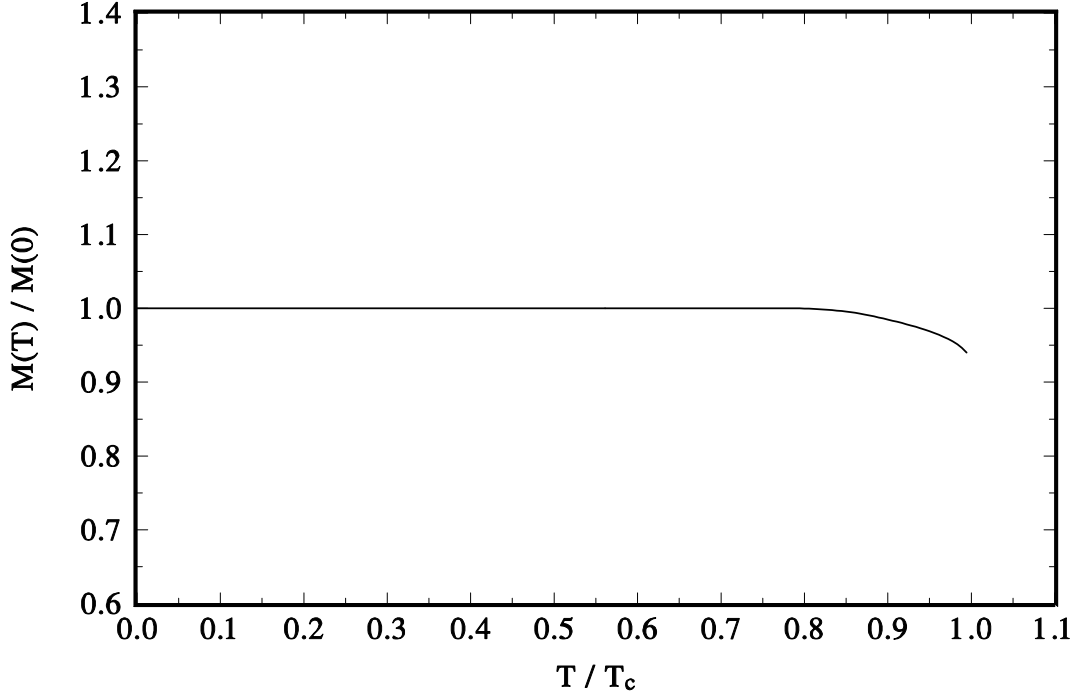}
\caption{The  ratio of the $\eta_c$ mass $M(T)/M(0)$  as a function of $T/T_c$ for $T_c =  180 \; \mbox{MeV}$. A similar result is found for the $\chi_c$, as well as for $T_c = 260 \; \mbox{MeV}$.}
\end{center}
\end{figure}

The thermal behaviour of the gluon condensate was obtained in \cite{latticeG}; to a good approximation this result can be written as

\begin{equation}
\left<\left<\frac{\alpha_s}{\pi}G^2\right>\right> = \left<\frac{\alpha_s}{\pi}G^2\right> \left[\theta(T^*-T) + \frac{1-\frac{T}{T_C}}{1-\frac{T^*}{T_C}}\theta(T-T^*) \right]
\end{equation}

where $T^*\approx 150$ MeV is the breakpoint temperature where the condensate begins  to decrease appreciably, and $T_C\approx 180$ MeV is the temperature at which $\left<\left<\frac{\alpha_s}{\pi}G^2\right>\right>_{T_C}=0$. It should be mentioned that this lattice QCD determination has rather large errors, so that this value of $T_c$ is affected by a large uncertainty. We have changed the value of $T_c$ in the wide range $T_c \simeq 180 - 260 \; \mbox{MeV}$, and found no qualitative change in the results, as discussed in the next section. This simple parametrization of the lattice data can be improved using a smooth function without changing the conclusions to be reached with Eq.(17). 
At T=0 the gluon condensate has been extracted from data on  $e^+ e^-$ annihilation, and $\tau$-lepton hadronic decays \cite{G2}:
$\langle \frac{\alpha_s}{\pi}G^2\rangle = (0.05 \pm 0.02) \mbox{GeV}^4$.
We do not consider a gluonic twist-two term in the OPE \cite{MOLEE} as it is negligible, as in the case of vector (J/$\psi$) charmonium \cite{JPSI}. In fact, after calculating the Wilson coefficients corresponding to the (pseudo) scalar correlator, the contribution from this term is roughly one order of magnitude smaller than the scalar gluon condensate at $T=T_c$.\\ 

\begin{figure}[ht]
\begin{center}
\includegraphics[width=\columnwidth]{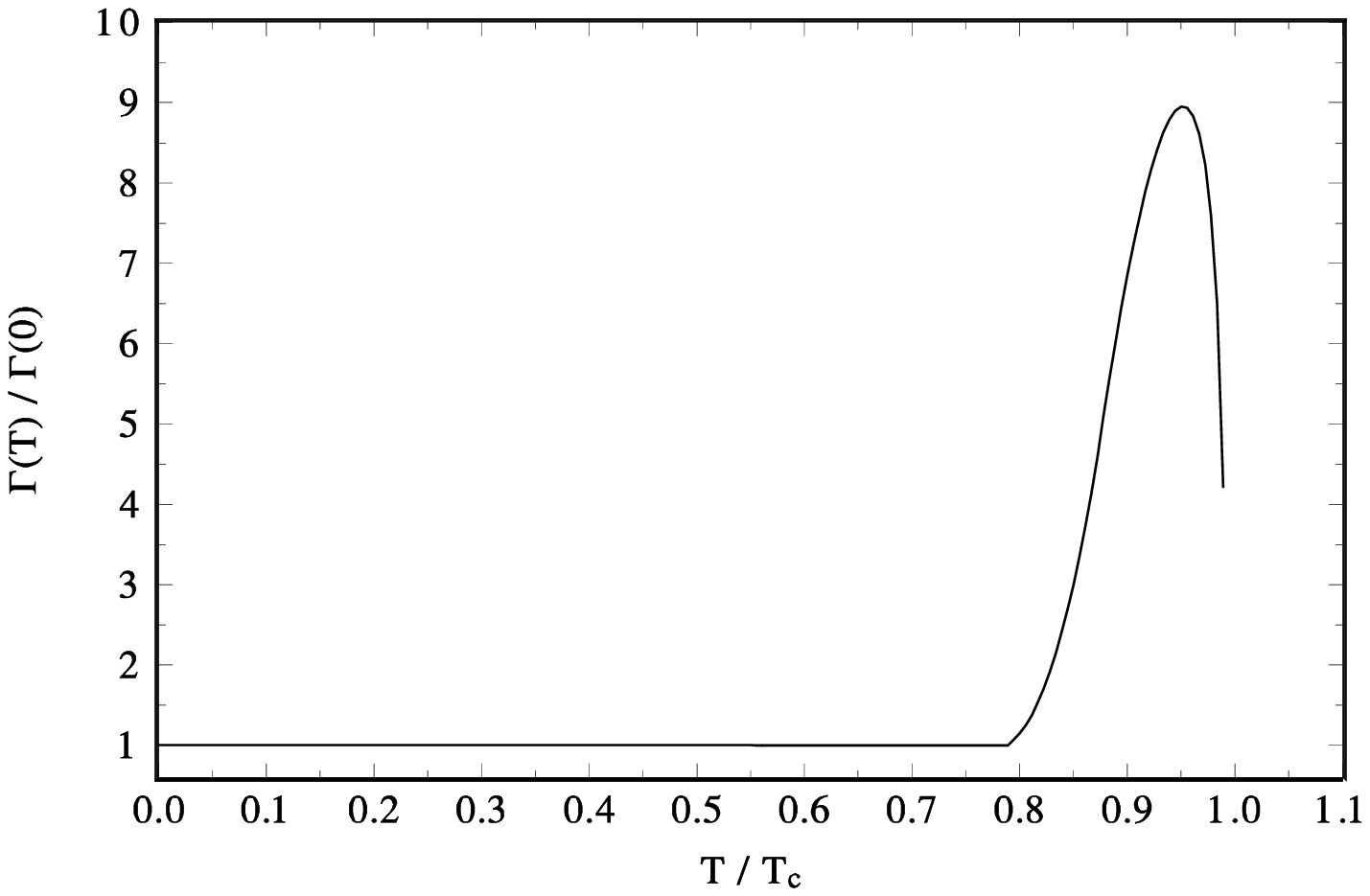}
\caption{The  ratio of the $\eta_c$ width  $\Gamma(T)/\Gamma(0)$  as a function of $T/T_c$ corresponding to $T_c = 180 \; \mbox{MeV}$. For $T_c = 260 \; \mbox{MeV}$ the break-point is at $T/T_c \simeq  0.55$, instead of $T/T_c \simeq 0.80$, so that the curve becomes roughly twice as broad.}
\end{center}
\end{figure}

Turning to the hadronic representation of the (pseudo)scalar correlator, we follow the standard procedure and  parametrize it in terms of the ground state resonance, i.e. the $\eta_c$ ($\chi_c$), followed   by a continuum given by PQCD after a threshold $s_0 > M^2$, where $M$ is the hadron mass. This ansatz is an even  better approximation at finite temperature, as $s_0(T)$ decreases monotonically with increasing $T$ in all systems analyzed so far. Considering first the zero width approximation, the hadronic spectral function is given by

\begin{eqnarray}
\frac{1}{\pi}\, Im \,\Pi(s,T)|_{HAD} &=&  \frac{1}{\pi} Im \,\Pi(s,T)|_{RES}\; \theta(s_0 - s) +  \frac{1}{\pi} Im \,\Pi(s,T)|_{PQCD} \;\theta(s - s_0)
\nonumber \\ [.3cm]
&=&  \, f^2(T) \, M^2(T) \, \delta(s - M^2(T)) \,+  \frac{1}{\pi} Im \,\Pi(s,T)^{a} \;\theta(s - s_0)\;,
\end{eqnarray}

where $s \equiv q^2 = \omega^2 - \mathbf{|q|}^2$, and the leptonic decay constant is defined as 

\begin{equation}
<0| J_{(5)}(0) | H(k)> = f\; M^2 \;,
\end{equation}

where H(k) stands for $\eta_c$ ($\chi_c$). Next, considering a finite (total) width  the following replacement will be understood

\begin{equation}
\delta(s- M^2(T)) \Longrightarrow const \; \frac{1}{(s-M^2(T))^2 + M^2(T) \Gamma^2(T)}\; ,
\end{equation}

where the constant is fixed by requiring equality of areas, e.g. if the integration is in the interval $(0 -\infty)$ then $ const =  M(T) \Gamma(T)/\pi$. To complete the hadronic parametrization one needs to consider the hadronic scattering term due to the current scattering off  heavy-light quark pseudoscalar mesons (D-mesons) in the thermal bath. In the case of the $\eta_c$, the pseudoscalar current couples to an odd number of pseudoscalars, so that the scattering term is effectively  a negligible higher order (one-loop) effect \cite{CAD1}. For the scalar case it is easy to show that the hadronic scattering spectral function vanishes in the rest frame of the thermal bath, as this term is not singular as $|\bf{q}| \rightarrow 0 $, and $\omega \rightarrow 0$.\\

\begin{figure}[ht]
\begin{center}
\includegraphics[width=\columnwidth]{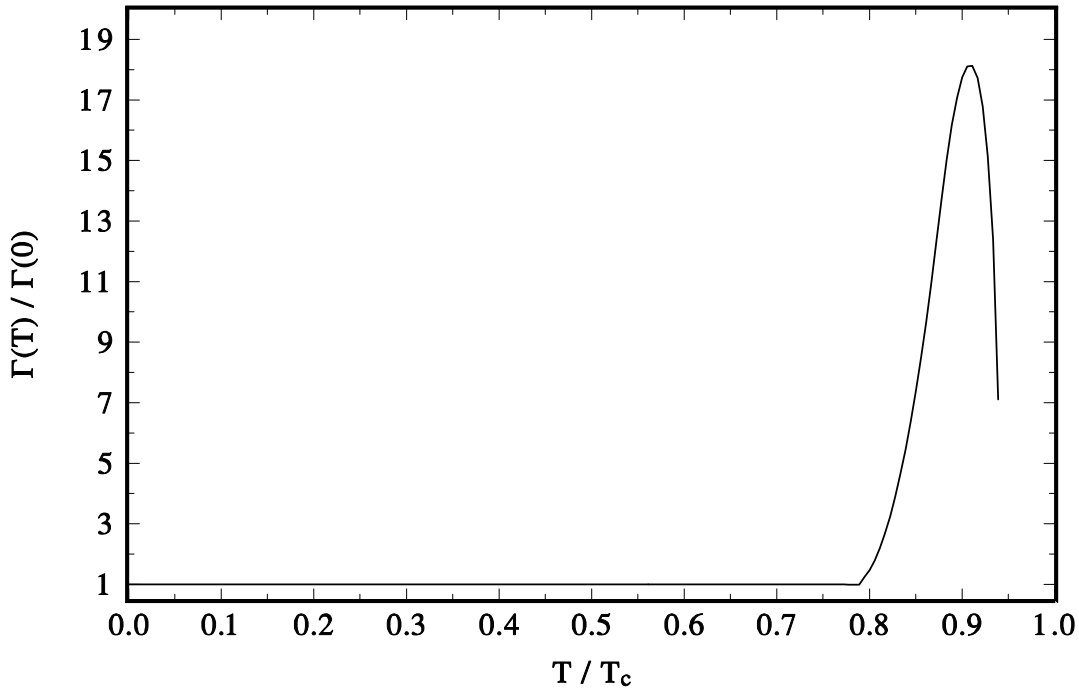}
\caption{The  ratio of the $\chi_c$ width  $\Gamma(T)/\Gamma(0)$  as a function of $T/T_c$ corresponding to $T_c = 180 \; \mbox{MeV}$. For $T_c = 260 \; \mbox{MeV}$ the break-point is at $T/T_c \simeq  0.55$, instead of $T/T_c \simeq 0.80$, so that the curve becomes roughly twice as broad..}
\end{center}
\end{figure}

\begin{figure}[ht]
\begin{center}
\includegraphics[width=\columnwidth]{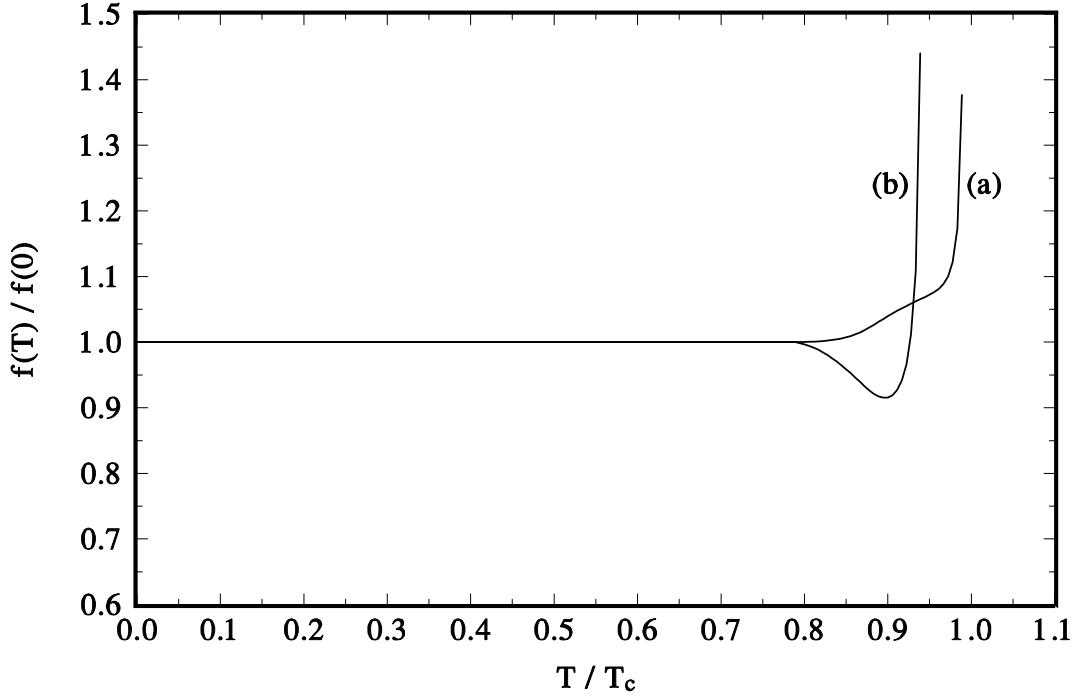}
\caption{The  ratio of the leptonic decay constant $f$  as a function of $T/T_c$ for the $\eta_c$ (curve (a)), and the $\chi_c$ (curve (b)) corresponding to $T_c = 180 \; \mbox{MeV}$. For $T_c = 260 \; \mbox{MeV}$ the break-point is at $T/T_c \simeq  0.55$, instead of $T/T_c \simeq 0.80$.}
\end{center}
\end{figure}

\section{ RESULTS}
We start the analysis at  $T=0$, and use the moments, Eq.(3), to reproduce the experimental values of the mass for an input width  of the $\eta_c$ and $\chi_c$ \cite{PDG}. The leptonic decay constants are not known experimentally, so that they will be predicted from the sum rules, having in mind that they are expected to be a few hundred MeV. Ultimately, their precise value is not important, as what matters is the behaviour of their ratio $f(T)/f(0)$. 
Beginning with the zero-width approximation it follows from Eq.(5) 
that

\begin{equation}
\frac{\varphi_1(Q^2)|_{HAD}}{\varphi_2(Q^2)|_{HAD}} = \frac{\varphi_2(Q^2)|_{HAD}}{\varphi_3(Q^2)|_{HAD}}\;,
\end{equation}

which turns out to hold with extremely good accuracy given the small widths of the $\eta_c$ and the $\chi_c$. Using Eq.(4) this leads to

\begin{equation}
\frac{\varphi_1(Q_0^2)|_{QCD}}{\varphi_2(Q_0^2)|_{QCD}} = \frac{\varphi_2(Q_0^2)|_{QCD}}{\varphi_3(Q_0^2)|_{QCD}}\;,
\end{equation} 

which depends only on the QCD information, involves the two unknowns $s_0$ and $Q_0^2$, and provides the first equation to determine this pair of parameters. It is very important to stress at this point that for consistency $s_0$ must be determined from the sum rules themselves, rather than importing it from some other analysis. The second equation can be Eq.(4) with e.g. $N=1$.
Following this procedure, and starting at $T=0$ we find for $\eta_c$ that $s_0(0) = 9.1\; \mbox{GeV}^2$,  $Q_0^2 = 0$, and $\Gamma(0) = \Gamma|_{EXP} =  27 \pm 3 \; \mbox{MeV}$ give $M(0) = 2.9 \; \mbox{GeV}$, to be compared with the experimental value $M(0)|_{EXP} = 2.98 \; \mbox{GeV}$. For the $\chi_c$ we find that $s_0(0) = 11.6 \; \mbox{GeV}^2$, $\Gamma = \Gamma|_{EXP} = 10.2 \pm 0.7 \;\mbox{MeV}$, and $Q_0^2 = 0$ yield $M(0) = 3.2 \; \mbox{GeV}$, to be compared with the experimental value $M(0)|_{EXP} = 3.41 \; \mbox{GeV}$. The leptonic decay constants are predicted to be $f = 183  \; \mbox{MeV}$ for $\eta_c$, and $f = 190  \; \mbox{MeV}$ for $\chi_c$, in agreement with expectations.
The  above results are fairly insensitive to changes in  $Q_0^2$ and $N$  in the range $Q_0^2 = 0 - 10\; \mbox{GeV}^2$, and $N = 1 - 6$ (for larger values of $N$ there is no solution for $s_0(0)$). A posteriori, and at finite temperature, varying $Q_0^2$ and N in that wide range leads to no appreciable changes in the results and conclusions.

Next, at finite $T$ Eq.(21) continues to hold with extreme accuracy even if the width were to grow substantially. In fact, for a tenfold increase in the width, Eq.(21) remains valid within a fraction of 1\%. Solving for $s_0(T)$ the results are shown in Fig.1 for $T_c = 180 \; \mbox{MeV}$, where curve (a) is for the pseudoscalar channel, and curve (b) for the scalar channel. The end point of each curve corresponds to $s_0(T_c) = 4 \,m_Q^2 \simeq 6.8 \; \mbox{GeV}^2$, i.e. the QCD threshold value which then makes the FESR integrals vanish. The break-point at which $s_0(T)$ starts to decrease occurs at $T/T_c \simeq \; 0.80$, for  $T_c = \; 180 \; \mbox{MeV}$ ($T/T_c \simeq \;0.55$ for $T_c = 260 \; \mbox{MeV}$). Since in the present case there is only one $T$ - dependent driving term, it is not possible to extend the analysis beyond the critical temperature. For the $J/\psi$ there are two driving terms, the PQCD scattering term proportional to $T^2$, and the gluon condensate defining a critical temperature \cite{JPSI}. This feature allowed the analysis to extend beyond $T = T_c$. Once again, $s_0(T)$ must be obtained from the QCD moments themselves, rather than using some temperature parametrization from some other application. Otherwise, results could be seriously inconsistent.
The result for the mass of the $\eta_c$ is shown in Fig.2; a similar behaviour is found for the $\chi_c$. This result confirms once again the view that the hadronic mass is not a meaningful {\it order} parameter, as it hardly changes with temperature; in some cases decreasing slightly, and in other cases increasing slightly with increasing $T$. Conceptually, there is no particular reason for the mass, being the position   of the spectral function pole on the real axis, to be a signal for deconfinement. In contrast,  the total width and the coupling do provide such a deconfinement signal, if the width increases and the coupling decreases with increasing temperature. In view of this result for the hadronic masses, we keep them constant in the sequel to simplify the analysis, and discuss later the consequences of allowing a slight $T$-dependence.
Results for the widths are shown in Fig.3 for $\eta_c$, and in Fig.4 for $\chi_c$. Qualitatively, they resemble the result for $J/\psi$. Quantitatively, the $\eta_c$ width is essentially independent of the temperature up to $T/T_c \simeq 0.80$ ($T/T_c \simeq \;0.55$ for $T_c = 260 \; \mbox{MeV}$), after which it increases substantially by a factor nine until  $T/T_c \simeq \;0.95$ ($T/T_c \simeq \;0.87$ for $T_c = 260 \; \mbox{MeV}$) where it decreases dramatically by more than a factor two at $T/T_c \simeq \;0.99$ ($T/T_c \simeq \;0.97$ for $T_c = 260 \; \mbox{MeV}$). Beyond this temperature there is no longer a solution to the FESR. The behaviour of the $\chi_c$ width is qualitatively similar to that of the  $\eta_c$. Quantitatively, though, the $\chi_c$ width increases by almost a factor eighteen until a slightly lower  $T/T_c \simeq \;0.90$ ($T/T_c \simeq \;0.75$ for $T_c = 260 \; \mbox{MeV}$), and then it decreases by a factor two at the end point $T/T_c \simeq \;0.94$ ($T/T_c \simeq \;0.85$ for $T_c = 260 \; \mbox{MeV}$). The behaviour of the leptonic decay constants is shown in Fig.5 for $\eta_c$ (curve (a)), and for $\chi_c$ (curve (b)). For the $\eta_c$, the temperature at which the coupling starts to increase dramatically is close to the temperature  at which the width begins to decrease substantially, i.e. $T/T_c \simeq \;0.95$ ($T/T_c \simeq \;0.97$ for $T_c = 260 \; \mbox{MeV}$). The end point is the same as for the width, i.e. $T/T_c \simeq \;0.99$ ($T/T_c \simeq \;0.97$ for $T_c = 260 \; \mbox{MeV}$).
A similar correlation is found for the $\chi_c$, i.e. the coupling reverses behaviour at  $T/T_c \simeq \; 0.90$ ($T/T_c \simeq \;0.75$ for $T_c = 260 \; \mbox{MeV}$), and the end point occurs at a  similar temperature as for the width, i.e  $T/T_c \simeq 0.94$ ($T/T_c \simeq \;0.85$ for $T_c = 260 \; \mbox{MeV}$). The origin of this correlation between the width and the coupling, and the fact that there is no longer a solution for the width and coupling (end point) below $T_c$ is easy to understand from Eqs.(4)-(6). For instance, in the zero-width approximation the square of the coupling is proportional to the right hand side of Eq.(6), i.e. to the sum of the PQCD integral and the non-perturbative (gluon condensate) moment, the latter being negative (see Eqs. (14) and (16)). At some  $T < T_c$ the monotonically decreasing PQCD integral becomes equal to the non-perturbative moment so that beyond this temperature $f^2 < 0$, and there is no longer a real solution. In the case of the widths the end point corresponds to the temperature beyond which they become negative. We comment in closing on the validity of the simplifying approximation $M(T) \simeq M(0)$. Allowing for the hadronic mass to change (decrease) near $T_c$ produces a slight reduction, together with a shift to the right, of the peak in Figs. 3 and 4, but preserving the overall qualitative behaviour of the widths. Something similar happens with the couplings, i.e. the temperature at which they start to increase substantially shifts slightly toward $T_c$.

\section{Conclusions}
\noindent
We have determined the temperature dependence of the hadronic parameters of charmonium in the (pseudo)scalar channels, $\eta_c$ and $\chi_c$, using Hilbert moment QCD sum rules. The results confirm the expectation that heavy-heavy quark systems behave differently from light-light and heavy-light quark hadrons, due to the absence of the light quark condensate in the OPE. Since in the space-like region there is no contribution to the PQCD nor to the hadronic (pseudo)scalar correlator, this leaves the gluon condensate as the leading term in the OPE, and thus solely responsible for the temperature behaviour of  $\eta_c$ and $\chi_c$. The continuum threshold $s_0(T)$ was found to follow closely the $T$-dependence of the gluon condensate. The leptonic decay constant and the width behave initially as in light and heavy-light hadrons, i.e. they both are initially constant, and after certain temperature the former decreases and the latter increases with increasing $T$. However, beyond a  certain temperature  this behaviour reverses, with the decay constant increasing and the width decreasing with $T$ up to some temperature just below $T_c$
beyond which the sum rules are no longer valid. In the (pseudo)scalar channel it is not possible to carry on the analysis beyond $T_c$, as  in the case of $J/\psi$, because the gluon condensate is the sole driver of the $T$ - dependence (there are no scattering terms as in the vector channel). 
Nevertheless,  the 
behaviour of the decay constant and the width of $\eta_c$ and $\chi_c$ near $T_c$  suggests their survival  beyond $T_c$, as found for the $J/\psi$ in this same framework \cite{JPSI}, and in agreement with lattice QCD results \cite{lattice}. Experimentally, a possible way of confirming the survival of charmonium states beyond $T_c$ would be to observe them in heavy ion collisions at high $p_T$ beyond a certain threshold value. 


\begin{thebibliography}{99}

\bibitem{REV} For a recent review see e.g. P. Colangelo and A. Khodjamirian, in: "At the Frontier of Particle Physics/ Handbook of QCD", M. Shifman, ed. (World Scientific, Singapore 2001), Vol. 3, 1495-1576.

\bibitem{BOCH} A.I. Bochkarev and M.E. Shaposnikov, Nucl. Phys. B {\bf 286}, 220 (1986).

\bibitem{CAD1} C.A. Dominguez and M. Loewe, Phys. Lett. B {\bf 233}, 201 (1989).
The (near) equality of the critical temperatures for chiral-symmetry restoration and
deconfinement was shown analytically in A. Barducci, R. Casalbuoni, S. De Curtis, R. Gatto
and G. Pettini, Phys. Lett. B {\bf 244}, 311 (1990). These authors used a result for the thermal quark condensate valid for $ 0 \leq T \leq T_c$, while the first reference only made use of the low-T expansion of chiral perturbation theory, obtaining somewhat different critical temperatures.

\bibitem{CAD2} C.A. Dominguez and M. Loewe, Physical Review  D {\bf 52},  3143 (1995).

\bibitem{VARIOUS} R.J. Furnstahl, T. Hatsuda and S.H. Lee,  Phys. Rev. D {\bf 42}, 1744 (1990);
C. Adami, T. Hatsuda and I. Zahed, Phys. Rev. D {\bf 43}, 921 (1991); C.A. Dominguez and M. Loewe, Z. Phys. C {\bf 49}, 423 (1991); {\it ibid.}
{\bf 51}, 69 (1991); {\it ibid.} {\bf 58}, 273 (1993);
C. Adami and I. Zahed, Phys. Rev. D {\bf 45}, 4312 (1992); T. Hatsuda, Y. Koike and S.-H. Lee, Phys. Rev. D {\bf 47},  1225 (1993); {\it ibid.} Nucl. Phys. B {\bf 394},  221 (1993); Y. Koike, Phys. Rev. D {\bf 48},  2313 (1993); C.A. Dominguez, M. Loewe and J.S. Rozowsky, Phys. Lett. B {\bf 335},  506 (1994);
 C.A. Dominguez, M. S. Fetea and M. Loewe, Phys. Lett.  B {\bf 387},  151 (1996); {\it ibid} B {\bf 406},  149 (1997); C.A. Dominguez, M. Loewe and  C. van Gend, Phys. Lett.  B {\bf 429},  64 (1998); {\it ibid} B {\bf 460},  442 (1999); C.A. Dominguez, and M. Loewe, Phys. Lett. B {\bf 481},  295 (2000).

\bibitem{HL} C.A. Dominguez, M. Loewe and J.C. Rojas, J. High Energy Phys. {\bf 0708},  040 (2007). 

\bibitem{JPSI} C. A. Dominguez, M. Loewe, J. C. Rojas, and Y. Zhang, Phys. Rev. D {\bf 81}, 014007 (2010).

\bibitem{DL95} C.A. Dominguez and M. Loewe, Phys. Rev. D {\bf 52}, 3143 (1995).

\bibitem{lattice} For recent results see e.g. H. Ohno, T. Umeda, and K. Kanaya, J. Phys. G {\bf 36}, 064027 (2009), and references therein.

\bibitem{RRYNP} L. J. Reinders, H. Rubinstein and S. Yazaki, Nucl. Phys. B {\bf 186}, 109 (1981).

\bibitem{mQ} T. Altherr and D. Seibert, Phys. Rev. C {\bf 49},  1684 (1994).

\bibitem{latticeG} G. Boyd and D. E. Miller, arXiv:hep-ph/9608482 (unpublished); D.E. Miller, arXiv:hep-ph/0008031 (unpublished).

\bibitem{G2}   R.A. Bertlmann, {\it et al.}, Z. Phys. C  {\bf 39},  231 (1988); C.A. Dominguez and J. Sola, Z. Phys. C {\bf 40},  63 (1988); C.A. Dominguez and K. Schilcher, J. High Energy Phys. {\bf 0701},  093 (2007).  
 
\bibitem{MOLEE} K. Morita and S.H. Lee, Phys. Rev. Lett. 100 (2008) 022301.

\bibitem{PDG}  K. Nakamura et al., Particle Data Group, J. Phys. G {\bf 37}, 075021 (2010).




\end{thebibliography}
\end{document}